\documentclass[aps,pre,twocolumn,amsmath,amssymb,superscriptaddress,longbibliography,showpacs,floatfix]{revtex4-2}
\usepackage{graphicx}
\begin{document}

\title{Large clusters in a correlated percolation model}
\author{Raz \surname{Halifa Levi}}\email{razhalifa@gmail.com}
\affiliation{The Faculty of Engineering,
Tel Aviv University, Tel Aviv 6997801, Israel}
\author{Yacov Kantor}
\affiliation{School of Physics and
Astronomy, Tel Aviv University, Tel Aviv 6997801, Israel}

\date{\today}

\begin{abstract}
We consider a correlated site percolation problem on a cubic lattice of  size $L^3$,
with $16\le L\le 512$. The sites of an initially full lattice are removed by a random
walk of ${\cal N}=uL^3$
steps. When the parameter $u$ crosses a threshold $u_c=3.15$, a large system transitions
between percolating and non-percolating states. We study the $L$-dependence of the
mean mass (number of sites) $M_r$ of the $r$th largest cluster, as well as $r$-dependence
of $M_r$ for various system sizes $L$ at $u_c$. We demonstrate that $M_r\sim L^{5/2}/r^{5/6}$
for moderate or large $L$ and $r\gg 1$, and also conclude that for {\em any} $r$ the
fractal  dimensions  of the clusters are $5/2$.
\end{abstract}

\maketitle

\section{Random walk percolation}\label{sec:intro}

The simplest (Bernoulli) percolation model consists of a lattice with randomly and
independently occupied sites or bonds. It is well understood and has been
extensively studied~\cite{Stauffer91,Grimmett99,Saberi15}. Many of its features
and behaviors also apply to correlated percolation models.  In this paper, we
consider a problem, where an initially full three-dimensional cubic lattice of
linear size $L$ (in lattice constants) and volume $L^3$ has its sites removed
by an ${\cal N}$-step random walk (RW) on the lattice. The {\em length} of the
RW that starts at a random position is proportional to the {\em volume} of the
lattice, namely, ${\cal N}=uL^3$. Periodic boundary conditions are imposed on
the walk on a finite lattice, i.e., the walker exiting through one boundary of
the lattice reemerges on the opposite boundary. (Since, the typical squared
distance traveled by a RW is proportional to the squared number of steps, the
number of such exits and re-entries is large ($\sim uL$). The parameter $u$
controls the length of the RW and, thus, the fraction of unvisited sites. This
three-dimensional  problem as well as its $d$-dimensional generalizations
have been studied by several authors
\cite{Banavar85,Abete04,KK_PRE100,FK_PRE103,Chalhoub24,HK_PRE110} and the
critical points as well as the critical exponents near those points have been
investigated.

Percolation theories are usually studied on regular lattices, where the
objects of interest are groups of connected neighboring present sites or
bonds that form {\em clusters}. In this paper we consider RW percolation model
of sites {\em unvisited} by a RW. A cluster is {\em spanning} if it contains
an uninterrupted path between the opposing boundaries in a specific direction.
In finite systems of linear size $L$ the probability $\Pi(u,L)$ of the existence
of such a path gradually decreases with increasing $u$. For an infinite system
$\Pi(u,\infty)$ becomes a step-function $\Theta(u_c-u)$ jumping from 1 to 0 at
the {\em critical value} $u_c$, called percolation threshold. The mean spatial extent (linear size) of the
{\em finite} clusters is called the correlation length $\xi$. Close to $u_c$
it diverges as
\begin{equation}\label{eq:nu}
\xi\sim|u-u_c|^{-\nu}\ ,
\end{equation}
where $\nu$ is the correlation length exponent. In RW percolation the probabilities
of sites {\em not visited} by a RW have power law correlations. Values of the
exponents $\nu$ for such systems in $d$ dimensions are known theoretically
~\cite{Weinrib84}, and when applied to RW percolation~~\cite{KK_PRE100} give
\begin{equation}\label{eq:nu_RW}
\nu=2/(d-2),\ \ {\rm for}\ \ 3\le d\le 6\ ,
\end{equation}
i.e., $\nu=2$ in $d=3$.

The fraction of sites $P$ belonging to the infinite cluster for $u<u_c$, also called the
{\em strength} of the infinite cluster, is
\begin{equation}\label{eq:beta}
P\sim (u_c-u)^{\beta}.
\end{equation}
Numerical studies and theoretical arguments \cite{Feshanjerdi23,Chalhoub24,HK_PRE110}
indicate that $\beta=1$ for all $d$, and we will use this value in the paper.

In Sec.~\ref{sec:clusters} we derive the theoretical relation between the mean mass of a large
cluster and its rank, while in Sec.~\ref{sec:ranksize} we numerically verify the predicted
relation between cluster sizes and their ranks.
Summary Sec.~\ref{sec:summary} briefly discusses the results.

\section{Cluster ranks and sizes}\label{sec:clusters}

The existence of an infinite cluster in an infinite percolating system presents
an interesting problem. Strictly away from the threshold in the percolating
region an infinite cluster is present with probability 1. However, {\em exactly at}
the threshold such a cluster is nowhere to be found ($P=0$) since its fractal
dimension \cite{Stauffer91}
\begin{equation}\label{eq:df0}
d_{\rm f}=d-\beta/\nu
\end{equation}
is smaller than the space dimension $d$ and it occupies zero fraction of the space,
and in that sense it is ``absent." The fractal nature of the large clusters dictates
the relation between their mass (number of sites) and their linear extent, and plays
an important role in their statistics. For the RW percolation, we can use
Eq.~\eqref{eq:nu_RW} and $\beta=1$ in Eq.~\eqref{eq:df0} to arrive at
\begin{equation}\label{eq:df}
d_{\rm f}=(d+2)/2\ ,
\end{equation}
which for $d=3$ gives $d_{\rm f}=5/2$.

The geometry, the number and the nature of the infinite clusters
have been considered by numerous authors - see, e.g.,
Refs.~\cite{Newman81,Newman81a,Aizenman97,Hofstad04}. Kesten \cite{Kesten86} made the
concept of the {\em incipient infinite cluster} at the percolation threshold more
rigorous by introducing the conditioning and the limiting schemes, and proved the
``existence" of such a cluster for several alternative definitions. (A convenient overview
of the subject can be found in Ref.~\cite{Borgs01}.)  In finite systems there may be
several alternative candidates for the would-be-infinite clusters, such as the largest
cluster in the sample, or, alternatively, the spanning cluster. The size distributions of
the largest non-spanning and spanning clusters are different \cite{Sen01}. The spanning
probability at the threshold is finite, and is presumed to be a universal number that
depends on $d$ and the macroscopic characteristics of the system, such as its aspect ratio.
(For Bernoulli percolation see, e.g., Refs.~\cite{Langlands92, Langlands94,Cardy92,Acharyya98},
while for RW percolation, see Refs.~\cite{Abete04,KK_PRE100}.)

The attempt to study the properties of a single ``infinite cluster in an infinite system"
by concentrating either on the largest or the largest spanning cluster in a finite system
slightly distorts the statistics because the linear dimensions of such a cluster are
close to $L$, thus leaving permanent influences of the boundaries. A different,
numerically useful, approach to the problem was suggested by Jan {\em et al.}~\cite{Jan98}
who used the fact that in a finite system of linear size $L$ at the  percolation threshold
there are many very large clusters that are appreciably smaller than $L$. Each one of
them can be viewed as a potential candidate for the infinite cluster as $L$ increases. (See also
the Refs.~\cite{Margolina82,Jan99}.) They {\em ranked} the clusters in each configuration as the
largest ({\em rank} $r=1$), second largest ($r=2$), and so on, and studied both the
$L$-dependence and the $r$-dependence of the cluster sizes. That approach followed Mandelbrot's
economic arguments \cite{Mandelbrot97} regarding the statistics of very large companies.
The main ideas of that argument are explained in the following four paragraphs.

The entire problem can be viewed via cluster statistics in finite systems of linear
dimension $L$: In a $d$-dimensional system of $N$ sites with $N_s$ clusters
containing $s$ sites, it is customary to  define {\em cluster density per site}
$n_s\equiv N_s/N$. (Obviously, those definitions require averaging over a suitable
ensemble.) Cluster statistics provides an alternative view of the percolation
problem. In particular, at the threshold and for $s\gg 1$ in an infinite
system, the typical length scales or typical cluster sizes are absent and the
expression for cluster density has a simple power-law form~\cite{Stauffer91}
\begin{equation}\label{eq:tau}
n_s=As^{-\tau},
\end{equation}
where the exponent $\tau>2$, since $\sum_ssn_s$ must converge to a fraction of
sites occupied by finite clusters~\cite{Stauffer91}.  Exponents describing the cluster
statistics are related to the ``macroscopic" (``thermal") exponents, and in particular
the exponent $\tau$ is related to $\beta$ and $\nu$~\cite{Stauffer91}, and, using the
expression \eqref{eq:df0} for the fractal dimension, it can be written as
\begin{equation}\label{eq:taudf}
\tau=1+\frac{d}{d_{\rm f}} ,
\end{equation}
which, using Eq.~\eqref{eq:df} for RW percolation becomes  $\tau=1+2d/(d+2)$, and for
$d=3$, we have $\tau=11/5$. The expression \eqref{eq:tau} can be used to calculate
the {\em number} of clusters {\em per site} larger than $s$:
\begin{align}\label{eq:Q}
Q(s)&=\int_s^\infty n_s ds=(\tau-1)^{-1}As^{1-\tau} \nonumber \\
    &=(\tau-1)^{-1}As^{-d/d_{\rm f}},
\end{align}
where in the last line we use Eq.~\eqref{eq:taudf}.

While the exponent $\tau$ depends only on the universality class of the percolation problem
and space dimension, the definitions of $n_s$ and $s$ are specifically related to
the lattice structure and the definition of the connectivity between adjacent sites.
As a result, the coefficient $A$ in Eqs.~\eqref{eq:tau} and \eqref{eq:Q} is
non-universal and depends on the details of the percolation. It is possible to rephrase
the above relation in a {\em universal} form: Cardy and Ziff \cite{Cardy03} approached
this problem by considering the linear {\em size} $\ell$ (in units of {\em length}) of
each  cluster, such as the radius of circumscribing (hyper)sphere, or maximal extent in
a particular direction, or the radius of gyration of the cluster. (Their work concentrated
on $d=2$ but provided guidance for a general $d$.) They argued that the number of clusters
$Q^*(\ell)$ larger than $\ell$ per unit volume, i.e., divided by system size ${\cal L}^d$
(in units of volume), in the continuum limit, where system size ${\cal L}\to\infty$, while
the lattice constant  $a\to0$, must have the only dimensionally possible form $C/\ell^d$
with dimensionless universal constant $C$. Since the masses  of
clusters are related to their sizes as $\ell^{d_{\rm f}}$ with
non-universal coefficient, while the volume of a site is determined by the lattice
constant, the universal function $Q^*(\ell)$ can be related to $Q(s)$ by non-universal
prefactors leading to non-universal $A$ in Eqs.~\eqref{eq:tau} and \eqref{eq:Q}.
We will proceed with our derivation using the non-universal form in Eq.~\eqref{eq:Q}.

Away from the threshold, there is an additional exponent that is used to describe the
truncation of the power law in Eq.~\eqref{eq:tau} for clusters with linear dimensions
exceeding $\xi$. Such truncation can be described by a function $\tilde{h}(s/\xi^{d_{\rm f}})$,
that multiplies to power law in Eq.~\eqref{eq:tau}, and decays fast when its argument
exceeds unity. Even at the percolation threshold, but at finite $L$ that power law will
be truncated by a similar function $h(s/L^{d_{\rm f}})$. Nevertheless, the expression for
$Q(s)$ in Eq.~\eqref{eq:Q} remains valid, if $s\ll L^{d_{\rm f}}$. In such a situation, this
expression for $Q(s)$ can be used to calculate the mean number of clusters $\tilde{r}$
larger than some mass $\tilde{M}_{\tilde{r}}$ in a system of size $L^d$. It will be
\begin{equation}\label{eq:r}
\tilde{r}=L^dQ(\tilde{M}_{\tilde{r}})\approx(\tau-1)^{-1}AL^d \tilde{M}_{\tilde{r}}^{-d/d_{\rm f}}\ ,
\end{equation}
provided $\tilde{r}\gg1$.

The relation \eqref{eq:r} can be approximately
used in an inverted form which relates the mean mass $M_r$ of the $r$th largest
cluster to the {\em rank} $r$ of that cluster and to the system size $L$.
\begin{equation}\label{eq:Mr}
M_r\sim L^{d_{\rm f}}/r^{d_{\rm f}/d}\ ,
\end{equation}
where we omitted a dimensionless prefactor of order unity. This relation is expected to
be valid for $r\gg1$, as long as $M_r\gg 1$.

For Bernoulli percolation, in two dimensions the power-law dependence of $M_r$ on $r$
was tested by several authors \cite{Jan98,Sen99} in various ranges of $r$. In $d=3$ this
dependence was studied in Ref.~\cite{Jan98a}.
(Sometimes the measured effective exponent of $r$ slightly deviated from the predictions.)
Even earlier, Watanabe \cite{Watanabe96,*Watanabe96a} studied the $r$-dependence of $M_r$
but he related his result to Zipf's law (see, e.g., p.55 in Ref.~\cite{Takayasu90}), rather
than the known percolation exponents, and his
result differed from the prediction of Eq.~\eqref{eq:Mr}.

If we use $d_{\rm f}$ assumed in Eq.~\eqref{eq:df} for RW percolation, we find
\begin{equation}\label{eq:MrRW}
M_r\sim L^{(d+2)/2}/r^{(d+2)/2d}\ .
\end{equation}
In $d=3$ this relation becomes $M_r\sim L^{5/2}/r^{5/6}$.

\begin{figure}[t]
\includegraphics[width=9 truecm]{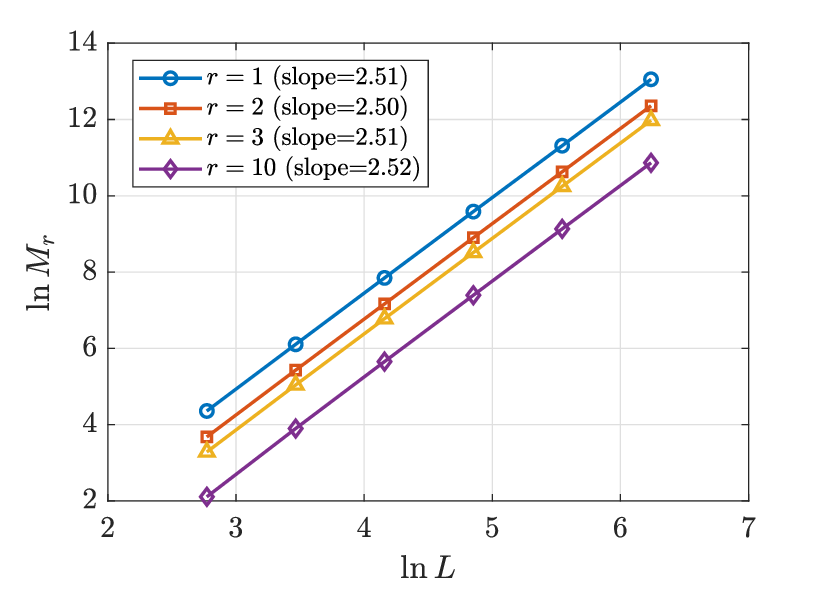}
\caption{Logarithmic plots of the mean mass $M_r$ of $r$th largest cluster at $u_c$,
averaged over $10^4$ configurations, as a
function of the system size $L$ ranging between $L=16$ and 512, for ranks (top to bottom)
$r=1$, 2, 3 and 10. Statistical errors of the points are smaller than the symbol sizes.
Slopes of the plots represent the fractal dimension $d_{\rm f}$.
}
\label{fig:MvsL}
\end{figure}

\section{Numerical results}\label{sec:ranksize}

In this work we consider realizations of the system at the percolation threshold
$u_c=3.15$. This value of the threshold was established in the previous works \cite{KK_PRE100,Chalhoub24,HK_PRE110} with accuracy better than $\pm 0.01$. For each
realization, the Hoshen-Kopelman algorithm \cite{Hoshen76} is used to identify {\em all}
the clusters. From the complete list, the clusters are ordered by ranks $r$, from the
largest ($r=1$) to the smallest of unit mass, and all the necessary quantities are
calculated. For each lattice size $L$ the procedure is repeated over $10^4$ samples,
and the averages for each $r$ are calculated. Large sample sizes produce about 1\%
accuracy for calculated averages. The linear system sizes are $L=16$, 32, \dots 512.

Figure \ref{fig:MvsL} depicts the system-size dependence of the masses of the clusters
of several ranks on a logarithmic scale. The slopes of the graphs are the fractal
dimensions $d_{\rm f}$ of the clusters of ranks $r$. Usually, only the largest cluster
($r=1$), which is a stand-in for the infinite cluster, is used to determine that dimension,
but in this example we clearly see the same behavior of clusters of various ranks.
Note that for each $L$ the mean mass of clusters of rank $r=10$ is approximately 10 times smaller
than the mass of the largest ($r=1$) cluster, and their linear sizes are about
2.5 times smaller.
The linear fits have errors due to the scatter of the points around the presumed straight
line that are smaller than 0.01. They are surprisingly good for fits including the modest $L=16$.
The range of $L$s is too small to detect a systematic changes of the slope due to increasing $L$.
The slopes vary between 2.50 and 2.52. We note that for $r=10$ the mean masses of the smallest
($L=16$) clusters are smaller than 10. The discreetness effects of such small clusters frequently
distort the measurement of the fractal dimension. Our estimates of $d_{\rm f}$ are close
to the $d_{\rm f}=5/2$ expected from Eq.~\eqref{eq:df}, assuming $\beta/\nu=1/2$.
This confirms our expectation that all large clusters have the same fractal dimension as
the infinite cluster.

\begin{figure}[t]
\includegraphics[width=9 truecm]{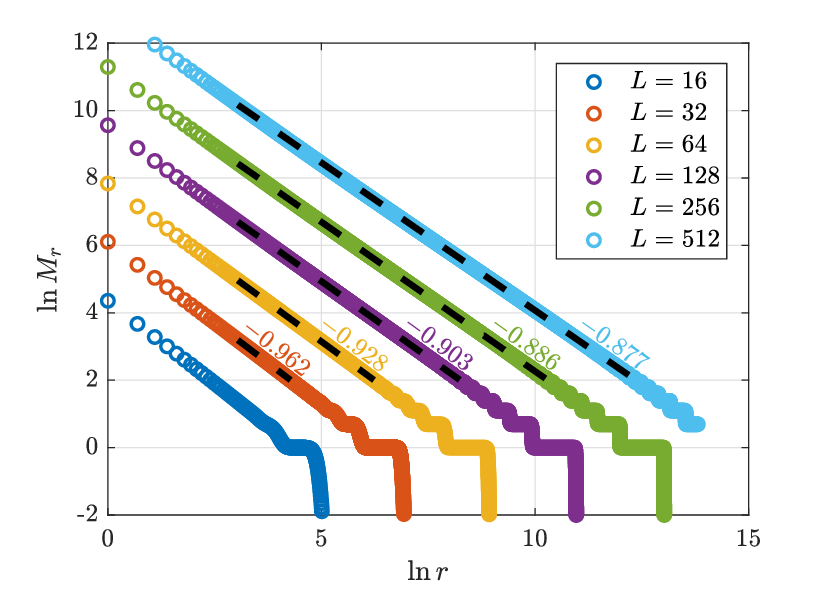}
\caption{Logarithmic plots of the mean mass $M_r$ of $r$th largest cluster as a
function of the rank $r$ for system sizes ranging (bottom-left to top-right) from $L=16$ to
$L=512$. Each data point has been averaged over an ensemble of $10^4$ configurations, and
statistical errors are significantly smaller than the symbol sizes.
The discreteness of specific values of $M_r$ becomes apparent in the steps at
the bottom of all graphs, when the ensemble averages cannot smooth-out the
discrete (integer) values of masses in each specific realization. For very
large $r$ there are no clusters ($M_r=0$) and the logarithm drops to $-\infty$.
Dashed lines are the linear fits in the selected range of $r$ and $M_r$ values,
and the numbers near them indicate their slopes. ($L=16$ has no valid range for the
required fit, but it has a rather straight segment of slope $\approx -1.0$.)
}
\label{fig:Mvsr}
\end{figure}

\begin{figure}[t]
\includegraphics[width=8.5 truecm]{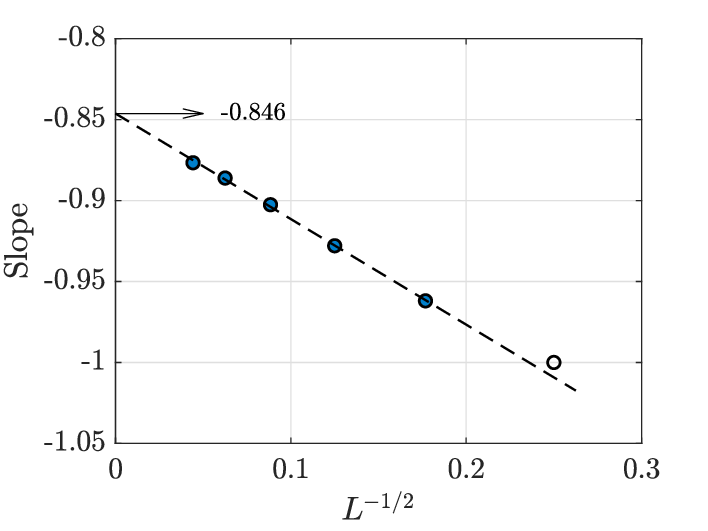}
\caption{Slopes of the fits in Fig.~\ref{fig:Mvsr} vs. $L^{-1/2}$, where $L$ is the
linear system size. Errors in the slopes are smaller than the symbol sizes. The open circle
represents the $L=16$ data which has no valid range for the fit. (See the caption of the
Fig.~\ref{fig:Mvsr}.) Only the slopes shown by the full circles represent the valid data,
and are used in the extrapolation $L\to\infty$. The arrow indicates the extrapolated value. }
\label{fig:extrapolate}
\end{figure}

Figure~\ref{fig:Mvsr} depicts on a logarithmic scale the numerical dependence of $M_r$
on $r$ for several values of $L$. The $r$-dependence is more complicated than the
$L$-dependence, since the exact relation depends on the cutoff function $h(s/L^{d_{\rm f}})$:
the power-law form of Eqs.~\eqref{eq:Mr} and \eqref{eq:MrRW} neglects the cutoff
function $h$ and is only valid for $r\gg 1$, i.e., for clusters with linear size significantly
smaller than $L$. This limitation, together with the requirement $M_r\gg 1$ severely
limits the range where Eqs.~\eqref{eq:Mr} and \eqref{eq:MrRW} are valid.
Since we expect that the linear size of the largest cluster $\ell_1\approx L$, we anticipate
$\ell_r/\ell_1\approx \ell_r/L=(M_r/M_1)^{1/d_{\rm f}}=1/r^{1/d}$, where we used
Eq.~\eqref{eq:Mr}. Thus, in $d=3$ we have $\ell_r/L\approx 1/r^{1/3}$. (For the purpose of
the estimate we use the power-law expression of $M_r$ dependence on $r$
even for small or moderate values of $r$, despite the fact that the exact power law
is no longer valid.) Thus, if we require that $\ell_r/L\lesssim 1/3$, then the value
of $r$ must exceed 20 or even 30.
On the other hand the power law dependence of $n_s$ in Eq.~\eqref{eq:tau} is valid only
when the cluster mass $s$ can be treated as a continuous variable.
The graphs in Fig.~\ref{fig:Mvsr} develop ``steps" when the value of $\ln M_r$ drops
below 2, i.e., $M_r\lesssim 8$.

Thus, the range of validity of  the power-law predictions is limited from both sides
of the graphs in Fig.~\ref{fig:Mvsr}. We made a ``generous" choice to limit the
range of the presumed power-law dependence in Fig.~\ref{fig:Mvsr} by $\ln M_r\ge 2$ and
$\ln r\ge 3$. With such a restriction the $L=16$ graph has no usable range for linear fit
(on a logarithmic scale). (Had we chosen ``stricter" limits, we would have further restricted
the usable range, and possibly eliminated the $L=32$ data.) Dashed lines in
Fig.~\ref{fig:Mvsr} for $L\ge 32$ show the range of the fits for each $L$, and the number
near those lines indicates the slope. The accuracy of the slopes of the presumed straight
lines is very high (better than 0.001). However we should keep in mind that the slope depends
on the choice of the fitting range, and can change as much as 0.01 for a different
choice of range.  Despite the range restrictions listed above, {\em all} the
graphs in Fig.~\ref{fig:Mvsr} have rather large straight stretches on a logarithmic scale.
Even $L=16$, that has no theoretically valid region, has a straight stretch of slope $-1.0$.
Interestingly enough, this apparent slope corresponds to Watanabe's
\cite{Watanabe96,*Watanabe96a} attribution of his result to Zipf's law.

From Eq.~\eqref{eq:MrRW} in $d=3$ the expected $r$-dependence of $M_r$ on a logarithmic
scale should have a slope of $-d_{\rm f}/d=-5/6\approx -0.833$. None of the straight-line fits in
Fig.~\ref{fig:Mvsr} has that slope. For small $L$ the slopes appreciably deviate
from the expected value, but for larger $L$s  the slopes gradually move towards the
expected value. In an infinite system, away from
the critical points some features deviate from the asymptotic behavior due to finite
correlation length $\xi$ (see Eq.~\eqref{eq:nu}). As $|u-u_c|\sim\xi^{-1/\nu}$ approaches
zero, we expect the deviations from the asymptotic behavior to vanish. For a finite
system at $u_c$ the $\xi\approx L$. Indeed, if we plot the slopes in Fig.~\ref{fig:Mvsr},
as a function of $L^{-1/\nu}=L^{-1/2}$, as shown in Fig.~\ref{fig:extrapolate}
they seem to have an almost linear dependence, which extrapolates to $\approx -0.846$ for
$L=\infty$. This extrapolation is rather close to the expected slope.

\section{Discussion}\label{sec:summary}

In this paper we demonstrated power law dependence of the mean mass of ranked clusters both on
system size $L$ and on the rank $r$ in the RW percolation model. The $L$-dependence of $M_r$ was accurate and
corresponded to the anticipated fractal dimension 5/2 already for moderate or even small values of
$L$. Note that this dependence was valid also for small values of $r$ that were
{\em outside} the range of power law dependence on $r$ predicted by Eq.~\eqref{eq:MrRW}.
On the other hand, the dependence on $r$ could be verified only by extrapolating the
observed powers to an infinite $L$. The disregarded cutoff function $h$ is causing these deviations.
Even for $L=512$ our results deviate by some 5\% from the expected result.
One can improve the results either from the knowledge of the function $h$, or by using
even larger $L$s. (Indeed, with an extra effort it might be possible to perform simulations
on system sizes significantly larger than 512.) Larger $L$s will also be needed to demonstrate
that even larger rank clusters (e.g., $r\sim 100$) have the same fractal dimension.

The linear size of the ranked clusters decreases only as $1/r^{1/3}$ and therefore there are many
large clusters of similar sizes. This suggests that the study of the properties of the
"infinite" cluster can have its statistics improved by considering a large number of clusters
of various ranks in each configuration.


\begin{thebibliography}{33}%
\makeatletter
\providecommand \@ifxundefined [1]{%
 \@ifx{#1\undefined}
}%
\providecommand \@ifnum [1]{%
 \ifnum #1\expandafter \@firstoftwo
 \else \expandafter \@secondoftwo
 \fi
}%
\providecommand \@ifx [1]{%
 \ifx #1\expandafter \@firstoftwo
 \else \expandafter \@secondoftwo
 \fi
}%
\providecommand \natexlab [1]{#1}%
\providecommand \enquote  [1]{``#1''}%
\providecommand \bibnamefont  [1]{#1}%
\providecommand \bibfnamefont [1]{#1}%
\providecommand \citenamefont [1]{#1}%
\providecommand \href@noop [0]{\@secondoftwo}%
\providecommand \href [0]{\begingroup \@sanitize@url \@href}%
\providecommand \@href[1]{\@@startlink{#1}\@@href}%
\providecommand \@@href[1]{\endgroup#1\@@endlink}%
\providecommand \@sanitize@url [0]{\catcode `\\12\catcode `\$12\catcode
  `\&12\catcode `\#12\catcode `\^12\catcode `\_12\catcode `\%12\relax}%
\providecommand \@@startlink[1]{}%
\providecommand \@@endlink[0]{}%
\providecommand \url  [0]{\begingroup\@sanitize@url \@url }%
\providecommand \@url [1]{\endgroup\@href {#1}{\urlprefix }}%
\providecommand \urlprefix  [0]{URL }%
\providecommand \Eprint [0]{\href }%
\providecommand \doibase [0]{https://doi.org/}%
\providecommand \selectlanguage [0]{\@gobble}%
\providecommand \bibinfo  [0]{\@secondoftwo}%
\providecommand \bibfield  [0]{\@secondoftwo}%
\providecommand \translation [1]{[#1]}%
\providecommand \BibitemOpen [0]{}%
\providecommand \bibitemStop [0]{}%
\providecommand \bibitemNoStop [0]{.\EOS\space}%
\providecommand \EOS [0]{\spacefactor3000\relax}%
\providecommand \BibitemShut  [1]{\csname bibitem#1\endcsname}%
\let\auto@bib@innerbib\@empty
\bibitem [{\citenamefont {Stauffer}\ and\ \citenamefont
  {Aharony}(1991)}]{Stauffer91}%
  \BibitemOpen
  \bibfield  {author} {\bibinfo {author} {\bibfnamefont {D.}~\bibnamefont
  {Stauffer}}\ and\ \bibinfo {author} {\bibfnamefont {A.}~\bibnamefont
  {Aharony}},\ }\href@noop {} {\emph {\bibinfo {title} {Introduction to
  Percolation Theory}}},\ \bibinfo {edition} {2nd}\ ed.\ (\bibinfo  {publisher}
  {Taylor and Francis},\ \bibinfo {address} {London, UK},\ \bibinfo {year}
  {1991})\BibitemShut {NoStop}%
\bibitem [{\citenamefont {Grimmett}(1999)}]{Grimmett99}%
  \BibitemOpen
  \bibfield  {author} {\bibinfo {author} {\bibfnamefont {G.}~\bibnamefont
  {Grimmett}},\ }\href@noop {} {\emph {\bibinfo {title} {Percolation}}},\
  \bibinfo {edition} {2nd}\ ed.\ (\bibinfo  {publisher} {Springer},\ \bibinfo
  {address} {Berlin},\ \bibinfo {year} {1999})\BibitemShut {NoStop}%
\bibitem [{\citenamefont {Saberi}(2015)}]{Saberi15}%
  \BibitemOpen
  \bibfield  {author} {\bibinfo {author} {\bibfnamefont {A.~A.}\ \bibnamefont
  {Saberi}},\ }\bibfield  {title} {\bibinfo {title} {Recent advances in
  percolation theory and its applications},\ }\href
  {http://dx.doi.org/10.1016/j.physrep.2015.03.003} {\bibfield  {journal}
  {\bibinfo  {journal} {Phys. Rep.}\ }\textbf {\bibinfo {volume} {578}},\
  \bibinfo {pages} {1} (\bibinfo {year} {2015})}\BibitemShut {NoStop}%
\bibitem [{\citenamefont {Banavar}\ \emph {et~al.}(1985)\citenamefont
  {Banavar}, \citenamefont {Muthukumar},\ and\ \citenamefont
  {Willemsen}}]{Banavar85}%
  \BibitemOpen
  \bibfield  {author} {\bibinfo {author} {\bibfnamefont {J.~R.}\ \bibnamefont
  {Banavar}}, \bibinfo {author} {\bibfnamefont {M.}~\bibnamefont
  {Muthukumar}},\ and\ \bibinfo {author} {\bibfnamefont {J.~F.}\ \bibnamefont
  {Willemsen}},\ }\bibfield  {title} {\bibinfo {title} {Fractal geometries in
  decay models},\ }\href@noop {} {\bibfield  {journal} {\bibinfo  {journal} {J.
  Phys. A: Math. Gen.}\ }\textbf {\bibinfo {volume} {18}},\ \bibinfo {pages}
  {61} (\bibinfo {year} {1985})}\BibitemShut {NoStop}%
\bibitem [{\citenamefont {Abete}\ \emph {et~al.}(2004)\citenamefont {Abete},
  \citenamefont {de~Candia}, \citenamefont {Lairez},\ and\ \citenamefont
  {Coniglio}}]{Abete04}%
  \BibitemOpen
  \bibfield  {author} {\bibinfo {author} {\bibfnamefont {T.}~\bibnamefont
  {Abete}}, \bibinfo {author} {\bibfnamefont {A.}~\bibnamefont {de~Candia}},
  \bibinfo {author} {\bibfnamefont {D.}~\bibnamefont {Lairez}},\ and\ \bibinfo
  {author} {\bibfnamefont {A.}~\bibnamefont {Coniglio}},\ }\bibfield  {title}
  {\bibinfo {title} {Percolation model for enzyme gel degradation},\ }\href
  {https://doi.org/10.1103/PhysRevLett.93.228301} {\bibfield  {journal}
  {\bibinfo  {journal} {Phys. Rev. Lett.}\ }\textbf {\bibinfo {volume} {93}},\
  \bibinfo {pages} {228301} (\bibinfo {year} {2004})}\BibitemShut {NoStop}%
\bibitem [{\citenamefont {Kantor}\ and\ \citenamefont
  {Kardar}(2019)}]{KK_PRE100}%
  \BibitemOpen
  \bibfield  {author} {\bibinfo {author} {\bibfnamefont {Y.}~\bibnamefont
  {Kantor}}\ and\ \bibinfo {author} {\bibfnamefont {M.}~\bibnamefont
  {Kardar}},\ }\bibfield  {title} {\bibinfo {title} {Percolation of sites not
  removed by a random walker in $d$ dimensions},\ }\href
  {https://doi.org/10.1103/PhysRevE.100.022125} {\bibfield  {journal} {\bibinfo
   {journal} {Phys. Rev. E}\ }\textbf {\bibinfo {volume} {100}},\ \bibinfo
  {pages} {022125} (\bibinfo {year} {2019})}\BibitemShut {NoStop}%
\bibitem [{\citenamefont {Federbush}\ and\ \citenamefont
  {Kantor}(2021)}]{FK_PRE103}%
  \BibitemOpen
  \bibfield  {author} {\bibinfo {author} {\bibfnamefont {A.}~\bibnamefont
  {Federbush}}\ and\ \bibinfo {author} {\bibfnamefont {Y.}~\bibnamefont
  {Kantor}},\ }\bibfield  {title} {\bibinfo {title} {Percolation perspective on
  sites not visited by a random walk in two dimensions},\ }\href@noop {}
  {\bibfield  {journal} {\bibinfo  {journal} {Phys. Rev. E}\ }\textbf {\bibinfo
  {volume} {103}},\ \bibinfo {pages} {032137} (\bibinfo {year}
  {2021})}\BibitemShut {NoStop}%
\bibitem [{\citenamefont {Chalhoub}\ \emph {et~al.}(2024)\citenamefont
  {Chalhoub}, \citenamefont {Drewitz}, \citenamefont {Pr\'evost},\ and\
  \citenamefont {Rodriguez}}]{Chalhoub24}%
  \BibitemOpen
  \bibfield  {author} {\bibinfo {author} {\bibfnamefont {C.}~\bibnamefont
  {Chalhoub}}, \bibinfo {author} {\bibfnamefont {A.}~\bibnamefont {Drewitz}},
  \bibinfo {author} {\bibfnamefont {A.}~\bibnamefont {Pr\'evost}},\ and\
  \bibinfo {author} {\bibfnamefont {P.-F.}\ \bibnamefont {Rodriguez}},\
  }\bibfield  {title} {\bibinfo {title} {Universality classes for percolation
  models with long-range correlations},\ }\href@noop {} {\bibfield  {journal}
  {\bibinfo  {journal} {ArXiv:2403.18787}\ } (\bibinfo {year}
  {2024})}\BibitemShut {NoStop}%
\bibitem [{\citenamefont {Halifa~Levi}\ and\ \citenamefont
  {Kantor}(2024)}]{HK_PRE110}%
  \BibitemOpen
  \bibfield  {author} {\bibinfo {author} {\bibfnamefont {R.}~\bibnamefont
  {Halifa~Levi}}\ and\ \bibinfo {author} {\bibfnamefont {Y.}~\bibnamefont
  {Kantor}},\ }\bibfield  {title} {\bibinfo {title} {Critical exponents of
  correlated percolation of sites not visited by a random walk},\ }\href
  {https://doi.org/10.1103/PhysRevE.110.024116} {\bibfield  {journal} {\bibinfo
   {journal} {Phys. Rev. E}\ }\textbf {\bibinfo {volume} {110}},\ \bibinfo
  {pages} {024116} (\bibinfo {year} {2024})}\BibitemShut {NoStop}%
\bibitem [{\citenamefont {Weinrib}(1984)}]{Weinrib84}%
  \BibitemOpen
  \bibfield  {author} {\bibinfo {author} {\bibfnamefont {A.}~\bibnamefont
  {Weinrib}},\ }\bibfield  {title} {\bibinfo {title} {Long-range correlated
  percolation},\ }\href@noop {} {\bibfield  {journal} {\bibinfo  {journal}
  {Phys. Rev. B}\ }\textbf {\bibinfo {volume} {29}},\ \bibinfo {pages} {387}
  (\bibinfo {year} {1984})}\BibitemShut {NoStop}%
\bibitem [{\citenamefont {Feshanjerdi}\ \emph {et~al.}(2023)\citenamefont
  {Feshanjerdi}, \citenamefont {Masoudi}, \citenamefont {Grassberger},\ and\
  \citenamefont {Ebrahimi}}]{Feshanjerdi23}%
  \BibitemOpen
  \bibfield  {author} {\bibinfo {author} {\bibfnamefont {M.}~\bibnamefont
  {Feshanjerdi}}, \bibinfo {author} {\bibfnamefont {A.~A.}\ \bibnamefont
  {Masoudi}}, \bibinfo {author} {\bibfnamefont {P.}~\bibnamefont
  {Grassberger}},\ and\ \bibinfo {author} {\bibfnamefont {M.}~\bibnamefont
  {Ebrahimi}},\ }\bibfield  {title} {\bibinfo {title} {Aftermath epidemics:
  {Percolation} on the sites visited by generalized random walk},\ }\href@noop
  {} {\bibfield  {journal} {\bibinfo  {journal} {Phys. Rev. E}\ }\textbf
  {\bibinfo {volume} {108}},\ \bibinfo {pages} {024312} (\bibinfo {year}
  {2023})}\BibitemShut {NoStop}%
\bibitem [{\citenamefont {Newman}\ and\ \citenamefont
  {Schulman}(1981{\natexlab{a}})}]{Newman81}%
  \BibitemOpen
  \bibfield  {author} {\bibinfo {author} {\bibfnamefont {C.~M.}\ \bibnamefont
  {Newman}}\ and\ \bibinfo {author} {\bibfnamefont {L.~S.}\ \bibnamefont
  {Schulman}},\ }\bibfield  {title} {\bibinfo {title} {Infinite cluster in
  percolation models},\ }\href@noop {} {\bibfield  {journal} {\bibinfo
  {journal} {J. Stat. Phys.}\ }\textbf {\bibinfo {volume} {26}},\ \bibinfo
  {pages} {613} (\bibinfo {year} {1981}{\natexlab{a}})}\BibitemShut {NoStop}%
\bibitem [{\citenamefont {Newman}\ and\ \citenamefont
  {Schulman}(1981{\natexlab{b}})}]{Newman81a}%
  \BibitemOpen
  \bibfield  {author} {\bibinfo {author} {\bibfnamefont {C.~M.}\ \bibnamefont
  {Newman}}\ and\ \bibinfo {author} {\bibfnamefont {L.~S.}\ \bibnamefont
  {Schulman}},\ }\bibfield  {title} {\bibinfo {title} {Number and density of
  percolating clusters},\ }\href@noop {} {\bibfield  {journal} {\bibinfo
  {journal} {J. Phys. A: Math. Gen.}\ }\textbf {\bibinfo {volume} {14}},\
  \bibinfo {pages} {1735} (\bibinfo {year} {1981}{\natexlab{b}})}\BibitemShut
  {NoStop}%
\bibitem [{\citenamefont {Aizenman}(1997)}]{Aizenman97}%
  \BibitemOpen
  \bibfield  {author} {\bibinfo {author} {\bibfnamefont {M.}~\bibnamefont
  {Aizenman}},\ }\bibfield  {title} {\bibinfo {title} {On the number of
  incipient spanning clusters},\ }\href@noop {} {\bibfield  {journal} {\bibinfo
   {journal} {Nucl. Phys. B}\ }\textbf {\bibinfo {volume} {485}},\ \bibinfo
  {pages} {551} (\bibinfo {year} {1997})}\BibitemShut {NoStop}%
\bibitem [{\citenamefont {van~der Hofstad}\ and\ \citenamefont
  {J{\'a}rai}(2004)}]{Hofstad04}%
  \BibitemOpen
  \bibfield  {author} {\bibinfo {author} {\bibfnamefont {R.}~\bibnamefont
  {van~der Hofstad}}\ and\ \bibinfo {author} {\bibfnamefont {A.~A.}\
  \bibnamefont {J{\'a}rai}},\ }\bibfield  {title} {\bibinfo {title} {The
  incipient infinite cluster for high-dimensional unoriented percolation},\
  }\href@noop {} {\bibfield  {journal} {\bibinfo  {journal} {J. Stat. Phys.}\
  }\textbf {\bibinfo {volume} {114}},\ \bibinfo {pages} {625} (\bibinfo {year}
  {2004})}\BibitemShut {NoStop}%
\bibitem [{\citenamefont {Kesten}(1986)}]{Kesten86}%
  \BibitemOpen
  \bibfield  {author} {\bibinfo {author} {\bibfnamefont {H.}~\bibnamefont
  {Kesten}},\ }\bibfield  {title} {\bibinfo {title} {The incipient infinite
  cluster in two-dimensional percolation},\ }\href@noop {} {\bibfield
  {journal} {\bibinfo  {journal} {Probab. Theory Relat. Fields}\ }\textbf
  {\bibinfo {volume} {73}},\ \bibinfo {pages} {369} (\bibinfo {year}
  {1986})}\BibitemShut {NoStop}%
\bibitem [{\citenamefont {Borgs}\ \emph {et~al.}(2001)\citenamefont {Borgs},
  \citenamefont {Chayes}, \citenamefont {Kesten},\ and\ \citenamefont
  {Spencer}}]{Borgs01}%
  \BibitemOpen
  \bibfield  {author} {\bibinfo {author} {\bibfnamefont {C.}~\bibnamefont
  {Borgs}}, \bibinfo {author} {\bibfnamefont {J.~T.}\ \bibnamefont {Chayes}},
  \bibinfo {author} {\bibfnamefont {H.}~\bibnamefont {Kesten}},\ and\ \bibinfo
  {author} {\bibfnamefont {J.}~\bibnamefont {Spencer}},\ }\bibfield  {title}
  {\bibinfo {title} {Finite-size scaling in percolation},\ }\href@noop {}
  {\bibfield  {journal} {\bibinfo  {journal} {Commun. Math. Phys.}\ }\textbf
  {\bibinfo {volume} {224}},\ \bibinfo {pages} {153} (\bibinfo {year}
  {2001})}\BibitemShut {NoStop}%
\bibitem [{\citenamefont {Sen}(2001)}]{Sen01}%
  \BibitemOpen
  \bibfield  {author} {\bibinfo {author} {\bibfnamefont {P.}~\bibnamefont
  {Sen}},\ }\bibfield  {title} {\bibinfo {title} {Nature of the largest cluster
  size distribution at the percolation threshold},\ }\href@noop {} {\bibfield
  {journal} {\bibinfo  {journal} {J. Phys. A: Math Gen.}\ }\textbf {\bibinfo
  {volume} {34}},\ \bibinfo {pages} {8477} (\bibinfo {year}
  {2001})}\BibitemShut {NoStop}%
\bibitem [{\citenamefont {Langlands}\ \emph {et~al.}(1992)\citenamefont
  {Langlands}, \citenamefont {Pichet}, \citenamefont {Pouliot},\ and\
  \citenamefont {Saint-Aubin}}]{Langlands92}%
  \BibitemOpen
  \bibfield  {author} {\bibinfo {author} {\bibfnamefont {R.~P.}\ \bibnamefont
  {Langlands}}, \bibinfo {author} {\bibfnamefont {C.}~\bibnamefont {Pichet}},
  \bibinfo {author} {\bibfnamefont {P.}~\bibnamefont {Pouliot}},\ and\ \bibinfo
  {author} {\bibfnamefont {Y.}~\bibnamefont {Saint-Aubin}},\ }\bibfield
  {title} {\bibinfo {title} {On the universality of crossing probabilities in
  two-dimensional percolation},\ }\href@noop {} {\bibfield  {journal} {\bibinfo
   {journal} {J. Stat. Phys.}\ }\textbf {\bibinfo {volume} {67}},\ \bibinfo
  {pages} {553} (\bibinfo {year} {1992})}\BibitemShut {NoStop}%
\bibitem [{\citenamefont {Langlands}\ \emph {et~al.}(1994)\citenamefont
  {Langlands}, \citenamefont {Pouliot},\ and\ \citenamefont
  {Saint-Aubin}}]{Langlands94}%
  \BibitemOpen
  \bibfield  {author} {\bibinfo {author} {\bibfnamefont {R.}~\bibnamefont
  {Langlands}}, \bibinfo {author} {\bibfnamefont {P.}~\bibnamefont {Pouliot}},\
  and\ \bibinfo {author} {\bibfnamefont {Y.}~\bibnamefont {Saint-Aubin}},\
  }\bibfield  {title} {\bibinfo {title} {Conformal invariance in
  two-dimensional percolation},\ }\href@noop {} {\bibfield  {journal} {\bibinfo
   {journal} {Bull. Amer. Phys. Soc.}\ }\textbf {\bibinfo {volume} {30}},\
  \bibinfo {pages} {1} (\bibinfo {year} {1994})}\BibitemShut {NoStop}%
\bibitem [{\citenamefont {Cardy}(1992)}]{Cardy92}%
  \BibitemOpen
  \bibfield  {author} {\bibinfo {author} {\bibfnamefont {J.~L.}\ \bibnamefont
  {Cardy}},\ }\bibfield  {title} {\bibinfo {title} {Critical percolation in
  finite confined-geometries},\ }\href
  {http://iopscience.iop.org/article/10.1088/0305-4470/25/4/009/meta}
  {\bibfield  {journal} {\bibinfo  {journal} {J. Phys. A: Math. Gen.}\ }\textbf
  {\bibinfo {volume} {25}},\ \bibinfo {pages} {L201} (\bibinfo {year}
  {1992})}\BibitemShut {NoStop}%
\bibitem [{\citenamefont {Acharyya}\ and\ \citenamefont
  {Stauffer}(1998)}]{Acharyya98}%
  \BibitemOpen
  \bibfield  {author} {\bibinfo {author} {\bibfnamefont {M.}~\bibnamefont
  {Acharyya}}\ and\ \bibinfo {author} {\bibfnamefont {D.}~\bibnamefont
  {Stauffer}},\ }\bibfield  {title} {\bibinfo {title} {Effects of boundary
  conditions on the critical spanning probability},\ }\href@noop {} {\bibfield
  {journal} {\bibinfo  {journal} {Int. J. Mod. Phys. C}\ }\textbf {\bibinfo
  {volume} {9}},\ \bibinfo {pages} {643} (\bibinfo {year} {1998})}\BibitemShut
  {NoStop}%
\bibitem [{\citenamefont {Jan}\ \emph {et~al.}(1998)\citenamefont {Jan},
  \citenamefont {Stauffer},\ and\ \citenamefont {Aharony}}]{Jan98}%
  \BibitemOpen
  \bibfield  {author} {\bibinfo {author} {\bibfnamefont {N.}~\bibnamefont
  {Jan}}, \bibinfo {author} {\bibfnamefont {D.}~\bibnamefont {Stauffer}},\ and\
  \bibinfo {author} {\bibfnamefont {A.}~\bibnamefont {Aharony}},\ }\bibfield
  {title} {\bibinfo {title} {An infinite number of effectively infinite
  clusters in critical percolation},\ }\href@noop {} {\bibfield  {journal}
  {\bibinfo  {journal} {J. Stat. Phys.}\ }\textbf {\bibinfo {volume} {92}},\
  \bibinfo {pages} {325} (\bibinfo {year} {1998})}\BibitemShut {NoStop}%
\bibitem [{\citenamefont {Margolina}\ \emph {et~al.}(1982)\citenamefont
  {Margolina}, \citenamefont {Hermann},\ and\ \citenamefont
  {Stauffer}}]{Margolina82}%
  \BibitemOpen
  \bibfield  {author} {\bibinfo {author} {\bibfnamefont {A.}~\bibnamefont
  {Margolina}}, \bibinfo {author} {\bibfnamefont {H.~J.}\ \bibnamefont
  {Hermann}},\ and\ \bibinfo {author} {\bibfnamefont {D.}~\bibnamefont
  {Stauffer}},\ }\bibfield  {title} {\bibinfo {title} {Size of largest and
  second largest cluster in random percolation},\ }\href@noop {} {\bibfield
  {journal} {\bibinfo  {journal} {Phys. Lett. A}\ }\textbf {\bibinfo {volume}
  {93}},\ \bibinfo {pages} {73} (\bibinfo {year} {1982})}\BibitemShut {NoStop}%
\bibitem [{\citenamefont {Jan}(1999)}]{Jan99}%
  \BibitemOpen
  \bibfield  {author} {\bibinfo {author} {\bibfnamefont {N.}~\bibnamefont
  {Jan}},\ }\bibfield  {title} {\bibinfo {title} {Large lattice random site
  percolation},\ }\href@noop {} {\bibfield  {journal} {\bibinfo  {journal}
  {Physica A}\ }\textbf {\bibinfo {volume} {266}},\ \bibinfo {pages} {72}
  (\bibinfo {year} {1999})}\BibitemShut {NoStop}%
\bibitem [{\citenamefont {Mandelbrot}(1997)}]{Mandelbrot97}%
  \BibitemOpen
  \bibfield  {author} {\bibinfo {author} {\bibfnamefont {B.~B.}\ \bibnamefont
  {Mandelbrot}},\ }\href@noop {} {\emph {\bibinfo {title} {Fractals and Scaling
  in Finance: Discontinuity, Concentration, Risk}}}\ (\bibinfo  {publisher}
  {Springer},\ \bibinfo {address} {Berlin},\ \bibinfo {year}
  {1997})\BibitemShut {NoStop}%
\bibitem [{\citenamefont {Cardy}\ and\ \citenamefont {Ziff}(2003)}]{Cardy03}%
  \BibitemOpen
  \bibfield  {author} {\bibinfo {author} {\bibfnamefont {J.}~\bibnamefont
  {Cardy}}\ and\ \bibinfo {author} {\bibfnamefont {R.~M.}\ \bibnamefont
  {Ziff}},\ }\bibfield  {title} {\bibinfo {title} {Exact results for the
  universal area distribution of clusters in percolation, {Ising}, and {Potts}
  models},\ }\href@noop {} {\bibfield  {journal} {\bibinfo  {journal} {J. Stat.
  Phys.}\ }\textbf {\bibinfo {volume} {110}},\ \bibinfo {pages} {1} (\bibinfo
  {year} {2003})}\BibitemShut {NoStop}%
\bibitem [{\citenamefont {Sen}(1999)}]{Sen99}%
  \BibitemOpen
  \bibfield  {author} {\bibinfo {author} {\bibfnamefont {P.}~\bibnamefont
  {Sen}},\ }\bibfield  {title} {\bibinfo {title} {On the universality of
  distribution of ranked clustermasses at critical percolation},\ }\href@noop
  {} {\bibfield  {journal} {\bibinfo  {journal} {J. Phys. A: Math Gen.}\
  }\textbf {\bibinfo {volume} {32}},\ \bibinfo {pages} {7673} (\bibinfo {year}
  {1999})}\BibitemShut {NoStop}%
\bibitem [{\citenamefont {Jan}\ and\ \citenamefont {Stauffer}(1998)}]{Jan98a}%
  \BibitemOpen
  \bibfield  {author} {\bibinfo {author} {\bibfnamefont {N.}~\bibnamefont
  {Jan}}\ and\ \bibinfo {author} {\bibfnamefont {D.}~\bibnamefont {Stauffer}},\
  }\bibfield  {title} {\bibinfo {title} {Random site percolation in three
  dimensions},\ }\href@noop {} {\bibfield  {journal} {\bibinfo  {journal} {Int.
  J. Mod. Phys. C}\ }\textbf {\bibinfo {volume} {9}},\ \bibinfo {pages} {341}
  (\bibinfo {year} {1998})}\BibitemShut {NoStop}%
\bibitem [{\citenamefont {Watanabe}(1996{\natexlab{a}})}]{Watanabe96}%
  \BibitemOpen
  \bibfield  {author} {\bibinfo {author} {\bibfnamefont {M.~S.}\ \bibnamefont
  {Watanabe}},\ }\bibfield  {title} {\bibinfo {title} {Zipf's law in
  percolation},\ }\href@noop {} {\bibfield  {journal} {\bibinfo  {journal}
  {Phys. Rev. E}\ }\textbf {\bibinfo {volume} {53}},\ \bibinfo {pages} {4187}
  (\bibinfo {year} {1996}{\natexlab{a}})}\BibitemShut {NoStop}%
\bibitem [{\citenamefont {Watanabe}(1996{\natexlab{b}})}]{Watanabe96a}%
  \BibitemOpen
  \bibfield  {author} {\bibinfo {author} {\bibfnamefont {M.~S.}\ \bibnamefont
  {Watanabe}},\ }\bibfield  {title} {\bibinfo {title} {Erratum: {Ziph's} law in
  percolation},\ }\href@noop {} {\bibfield  {journal} {\bibinfo  {journal}
  {Phys. Rev. E}\ }\textbf {\bibinfo {volume} {54}},\ \bibinfo {pages} {4483}
  (\bibinfo {year} {1996}{\natexlab{b}})}\BibitemShut {NoStop}%
\bibitem [{\citenamefont {Takayasu}(1990)}]{Takayasu90}%
  \BibitemOpen
  \bibfield  {author} {\bibinfo {author} {\bibfnamefont {H.}~\bibnamefont
  {Takayasu}},\ }\href@noop {} {\emph {\bibinfo {title} {Fractals in the
  Physical Sciences}}}\ (\bibinfo  {publisher} {Manchester University Press},\
  \bibinfo {address} {Manchester, UK},\ \bibinfo {year} {1990})\BibitemShut
  {NoStop}%
\bibitem [{\citenamefont {Hoshen}\ and\ \citenamefont
  {Kopelman}(1976)}]{Hoshen76}%
  \BibitemOpen
  \bibfield  {author} {\bibinfo {author} {\bibfnamefont {J.}~\bibnamefont
  {Hoshen}}\ and\ \bibinfo {author} {\bibfnamefont {R.}~\bibnamefont
  {Kopelman}},\ }\bibfield  {title} {\bibinfo {title} {Percolation and cluster
  distribution. {I.} {Cluster} multiple labeling technique and critical
  concentration algorithm},\ }\href {https://doi.org/10.1103/PhysRevB.14.3438}
  {\bibfield  {journal} {\bibinfo  {journal} {Phys. Rev. B}\ }\textbf {\bibinfo
  {volume} {14}},\ \bibinfo {pages} {3438} (\bibinfo {year}
  {1976})}\BibitemShut {NoStop}%
\end{thebibliography}
\end{document}